\documentclass[3p,times]{elsarticle}
\usepackage{ecrc}
\volume{00}
\firstpage{1}
\journalname{Nuclear Physics A}
\runauth{}
\jid{npa}
\jnltitlelogo{Nuclear Physics A}
\usepackage{amssymb}
\biboptions{square,comma,numbers,sort&compress}
\usepackage[figuresright]{rotating}
\newcommand{\pp}{pp}

\newcommand{\ppb}{p+Pb}
\newcommand{\dau}{d+Au}
\newcommand{\pau}{p+Au}
\newcommand{\snn}{\sqrt{s_{_{NN}}}}
\newcommand{\gev}{GeV/$c$}
\newcommand\pt{p_T}

\newcommand\dphi{\Delta\phi}
\newcommand\deta{\Delta\eta}
\newcommand\mean[1]{\langle#1\rangle}
\begin{document}
\begin{frontmatter}
\dochead{}
\title{Dihadron azimuthal correlations at large pseudo-rapidity difference in multiplicity-selected d+Au collisions by STAR}
\author{Fuqiang Wang (for the STAR Collaboration)}
\address{Department of Physics and Astronomy, Purdue University, West Lafayette, Indiana 47907, USA}
\begin{abstract}
The long-range pseudo-rapidity correlations observed in \pp\ and \ppb\ collisions at the LHC have raised extensive interest. The tantalizing observation of similar effect by PHENIX calls for careful examination of the RHIC \dau\ data. 
In this talk, we present dihadron correlations in multiplicity-selected \dau\ collisions by STAR, in both $\dphi$ and $\deta$ and from both TPC and FTPC. 
\end{abstract}
\end{frontmatter}

\section{Introduction}

A long-range pseudo-rapidity correlation is unexpectedly observed in \pp~\cite{CMSppRidge} and \ppb~\cite{CMSpPbRidge,ALICEpPbRidge} 
collisions at the LHC after a uniform background subtraction. It is called the ``ridge,'' in analogy to the similar phenomenon in heavy-ion collisions after subtraction of an elliptic flow background~\cite{PRL95}. 
The heavy-ion ridge has been attributed primarily to triangular anisotropy, resulting from initial geometry fluctuations and subsequent hydrodynamic evolution~\cite{AlverV3}. The question arises whether the \pp\ and \ppb\ ridges are of the similar origin. Indeed, hydrodynamic calculations with event-by-event geometry fluctuations can describe the observed ridge in \pp\ and \ppb\ collisions~\cite{Bozek:2010pb}. 
However, other physics mechanisms are also possible, such as the color glass condensate where two-gluon density is relatively enhanced at small $\dphi$ over wide $\deta$~\cite{Dumitru:2010iy}. 


Furthermore, a back-to-back double ridge is observed by subtracting dihadron correlations in peripheral \ppb\ from that in central collisions~\cite{ALICEpPbRidge}. 
If jet correlations--which dominate the away-side correlations--are equal between peripheral and central collisions, then the observed double ridge would be an indication of non-jet physics. Jet correlations are due to hard scattering and not expected to differ over \ppb\ collision {\em centrality} except for nuclear $k_T$ effects. However, centrality is often defined by measured multiplicity, which can bias events with varying magnitudes of jet correlations. 

PHENIX analyzed their \dau\ data using the same ``central $-$ peripheral'' technique in their acceptance of $|\deta|<0.7$~\cite{PHENIXdAuRidge}. A double ridge was observed in the ``central $-$ peripheral'' dihadron correlation difference. 
In fact, two-particle correlations in \pp\ and \dau\ collisions are not identical and a subtle difference has been observed previously at RHIC~\cite{STARflow05}.
While it is an open question how much jet contribution remains in the PHENIX result in their limited acceptance, the complementarity between LHC and RHIC can be potentially powerful to distinguish the proposed ridge production mechanisms. 

STAR, with its large acceptance, is suitable to investigate centrality biases to dihadron correlations. 
In this talk, we present STAR results of dihadron correlations in \dau\ collisions as a function of multiplicity, with the large acceptance of $|\deta|<2$ of the STAR Time Projection Chamber (TPC). We also present dihadron correlations using STAR's mid-rapidity TPC and forward TPC (FTPC), with a $|\deta|$ coverage of 1.8-4.8. We examine the $\deta$ dependence of the correlations as well as the difference in the correlations between central and peripheral collisions. We discuss our results in the context of the LHC and PHENIX data.



\section{Data Sample and Data Analysis}
The data presented here were taken during the \dau\ run in 2003~\cite{Levente09}. The coincidence of the signals from the Zero Degree Calorimeters (ZDC) and the Beam-Beam Counters (BBC) selects minimum-bias (MB) \dau\ collisions. Events used in this analysis are required to have a primary vertex position $|z_{\rm vtx}|<30$~cm from the TPC center. A total of approximately 10 million events were used. TPC(FTPC) tracks are required to have at least 25(5) out of maximum possible 45(10) hits and a distance of closest approach to the primary vertex within 3~cm.

Three quantities were used to define \dau\ centrality: charged particle multiplicity within $|\eta|<1$ measured by the TPC, charged particle multiplicity within $-3.8<\eta<-1.8$ measured by the FTPC in the Au-beam direction (FTPC-Au)~\cite{Levente09}, and neutral energy measured in the ZDC of the Au-beam direction (ZDC-Au). 
Positive correlations are observed between these centrality measures but the correlations are quite broad. The same percentile centralities defined by different centrality measures correspond to significantly different event samples of \dau\ collisions.

We also cut on the ZDC-d signal to select the single neutron peak to enhance \pau\ collisions. The \pau\ results are similar to the \dau\ results.

Two sets of dihadron correlations are analyzed: TPC-TPC correlations where the trigger and associated particles are both from the TPC within $|\eta|<1$, and TPC-FTPC correlations where the trigger particle is from the TPC and the associated particle is from either the FTPC-Au within $-3.8<\eta<-2.8$ or the FPTC-d within $2.8<\eta<3.8$. The $\pt$ ranges of trigger and associated particles reported here are both $1<\pt<3$~\gev. The results are corrected for the tracking efficiency of $85\%\pm4\%$~(syst.) and $70\%\pm4\%$~(syst.) for associated particles in the TPC and FTPC, respectively. The correlations are normalized per trigger particle.

The two-particle acceptance correction is obtained from the mixed-events technique. The mixed events are required to be within 5~cm in $z_{\rm vtx}$, with the same multiplicity (for the TPC and FTPC-Au centrality measures) or within ZDC-Au ADC-Sum 10-size bins (for the ZDC-Au centrality measure). The mixed-events acceptance is normalized to 100\% at $\deta=0$ for TPC-TPC correlations and at $\deta=\pm3$ for TPC-FTPC correlations.

Two approaches are taken to analyze the correlation functions. One is to analyze the correlated yields after subtracting a uniform combinatorial background. The background is normalized by the Zero-Yield-At-Minimum (ZYAM) assumption~\cite{ZYAM}. The other is to decompose the correlation functions into a Fourier series and study the Fourier coefficients. No background subtraction is required for this method.

Systematic uncertainties are assessed by varying the ZYAM normalization $\dphi$ range from the default of 0.4 to 0.2 and 0.6 radian. 
The correlated yield has an additional 5\% systematic uncertainty due to the uncertainty in the tracking efficiency. For the Fourier coefficients, the systematic uncertainties are expected to be small compared to statistical uncertainties, but a thorough study of systematic uncertainties has not been done yet.

\section{Results and Discussions}
\begin{figure}
\begin{center}
\includegraphics[width=0.325\textwidth]{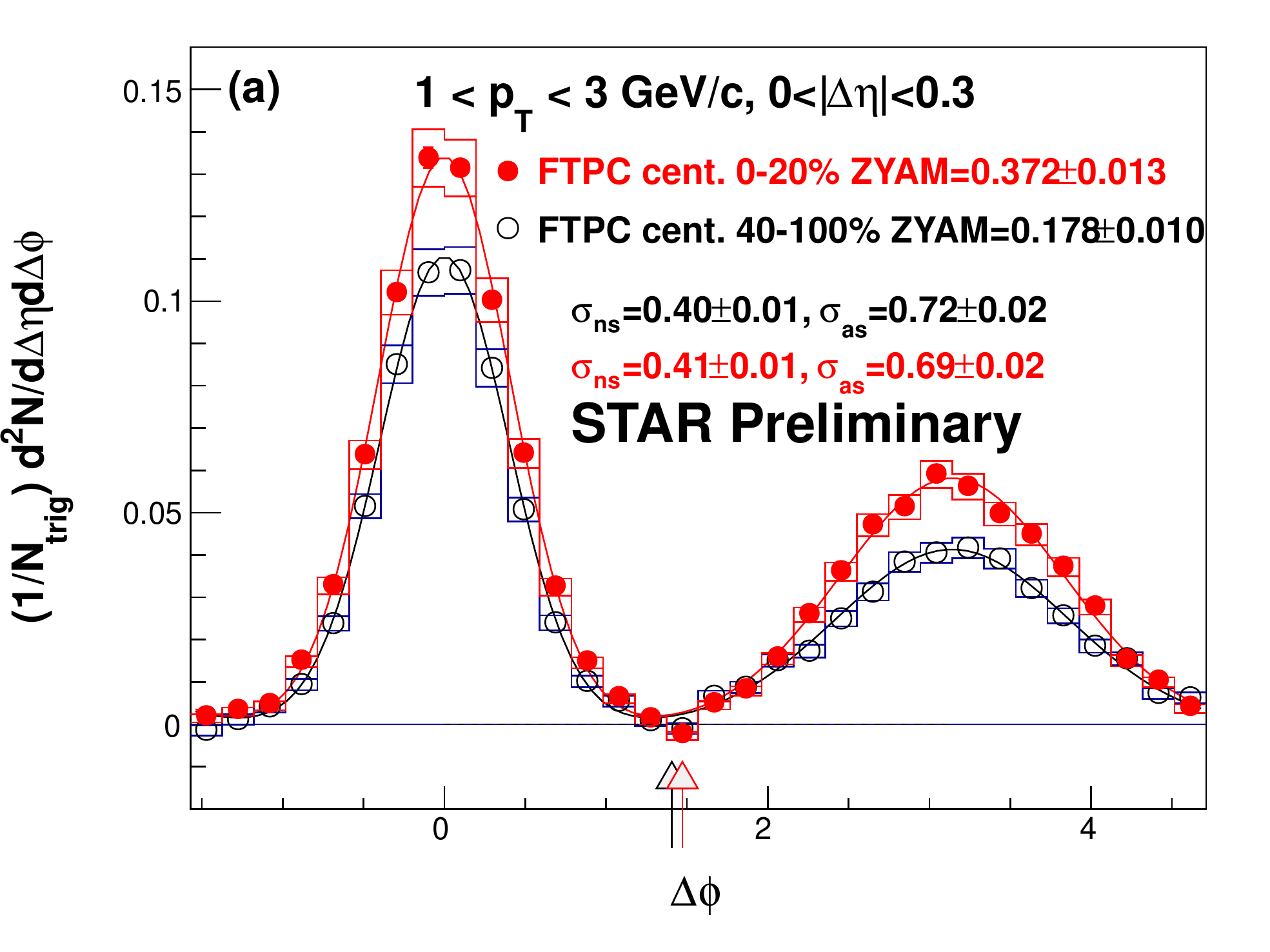}
\includegraphics[width=0.325\textwidth]{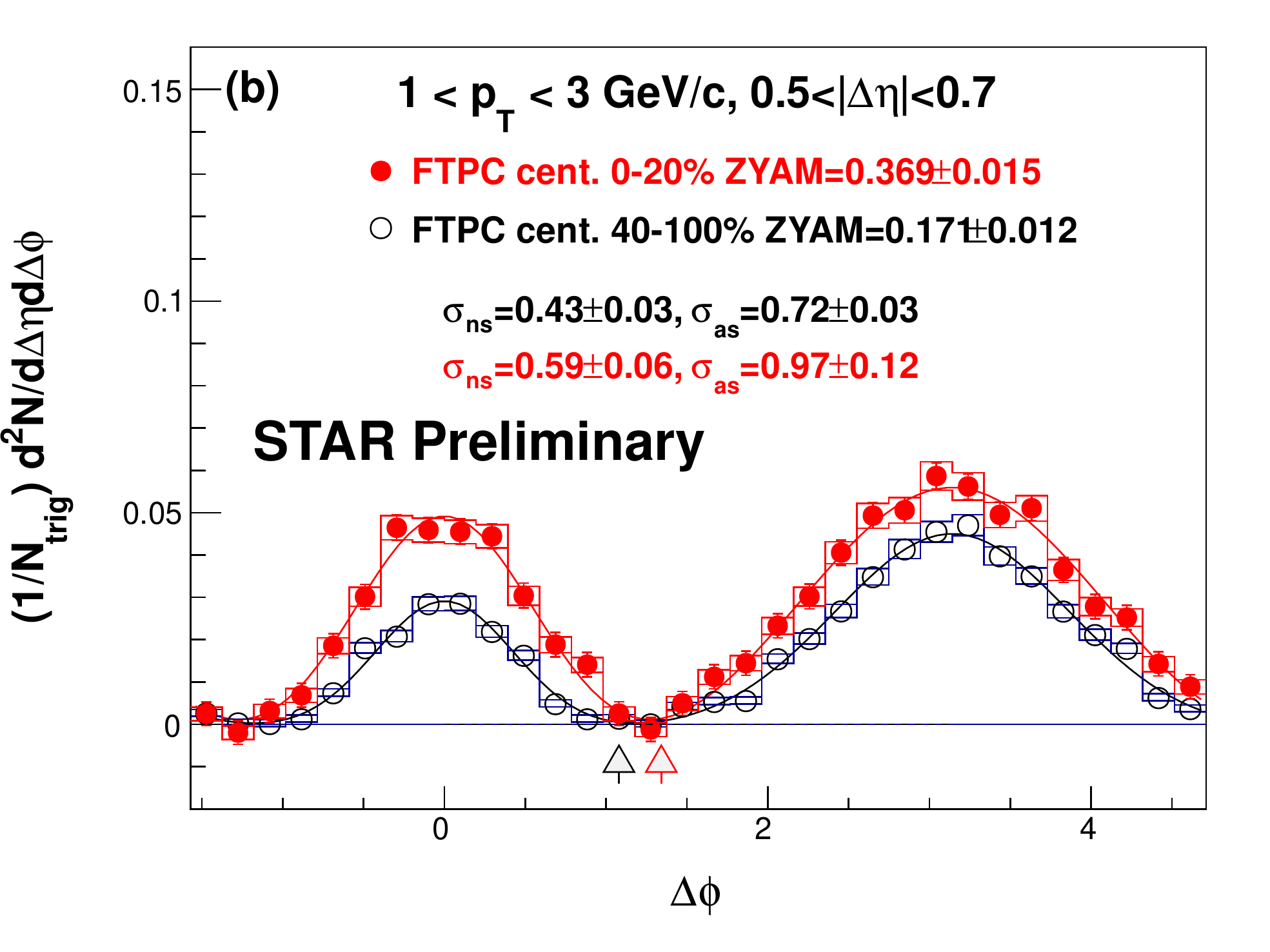}
\includegraphics[width=0.325\textwidth]{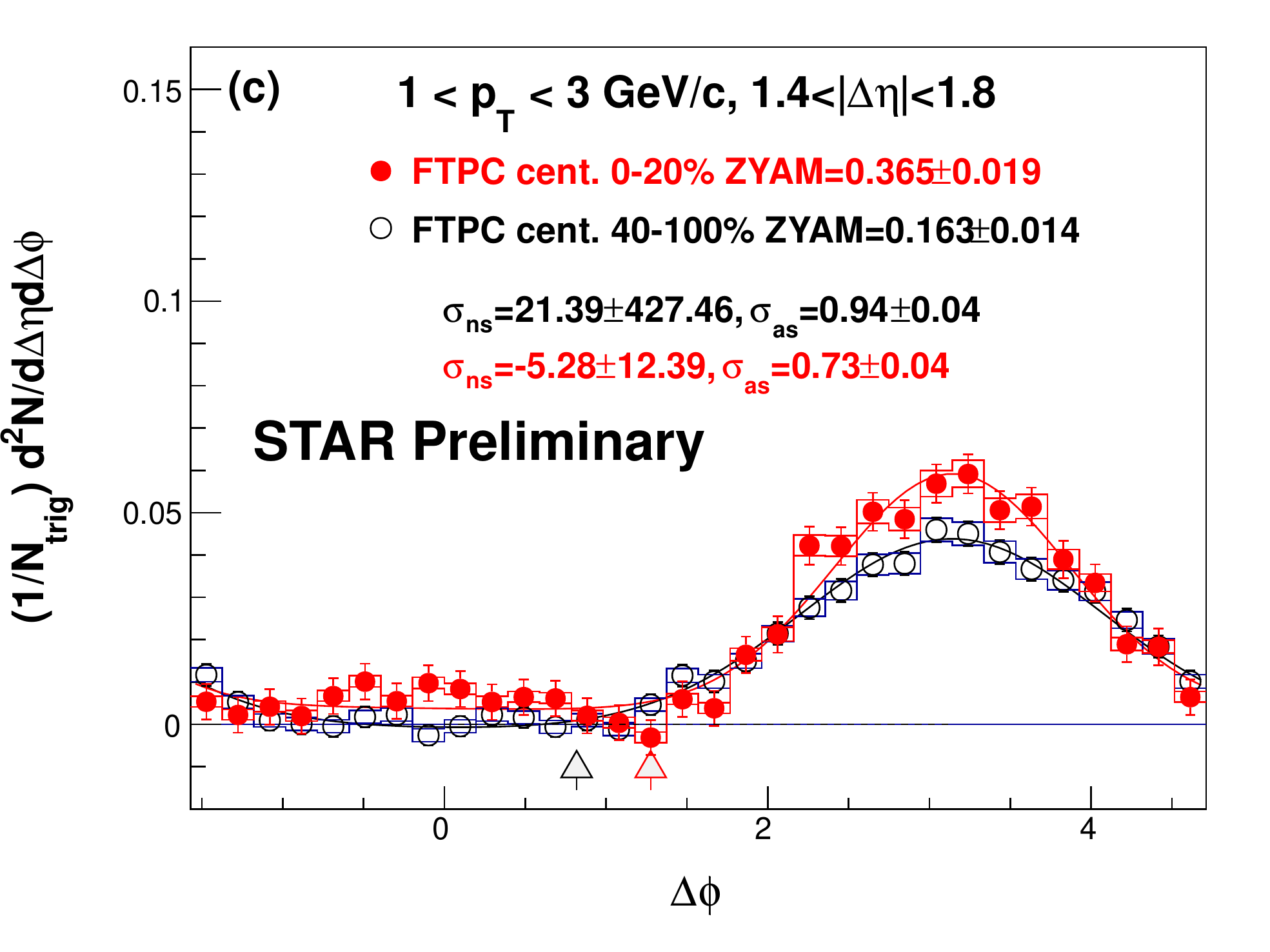}
\end{center}
\caption{Dihadron $\dphi$ correlations in three ranges of $|\deta|$ in peripheral (black) and central (red) \dau\ collisions. Trigger and associated particles are both from TPC ($|\eta|<1$) and $1<\pt<3$~\gev. Centrality is determined by FTPC-Au ($-3.8<\eta<-2.8$) multiplicity. The arrows indicate ZYAM normalization locations. Error bars are statistical and boxes indicate systematic uncertainties.}
\label{fig:dphi}
\begin{center}
\includegraphics[width=0.325\textwidth]{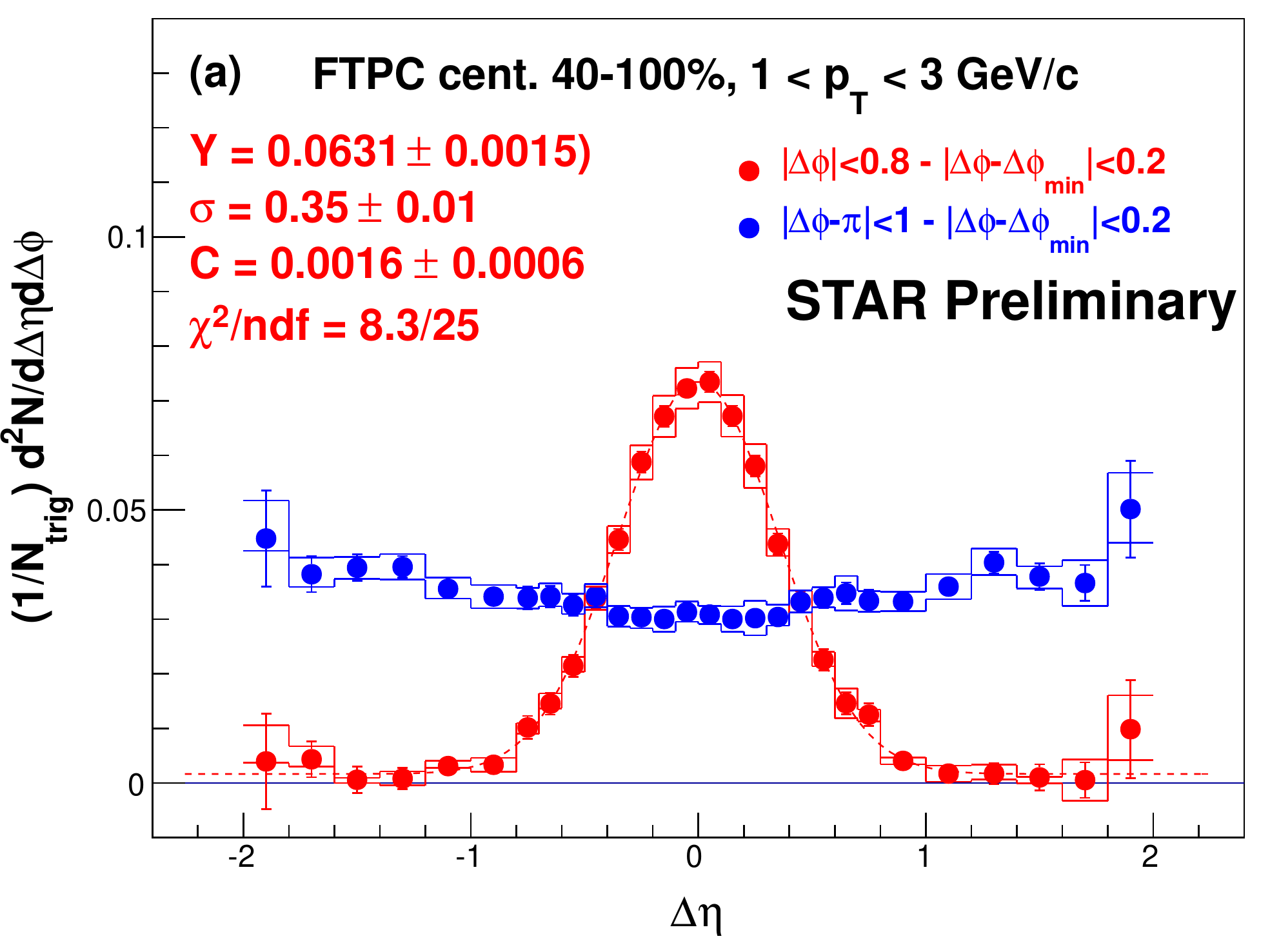}
\includegraphics[width=0.325\textwidth]{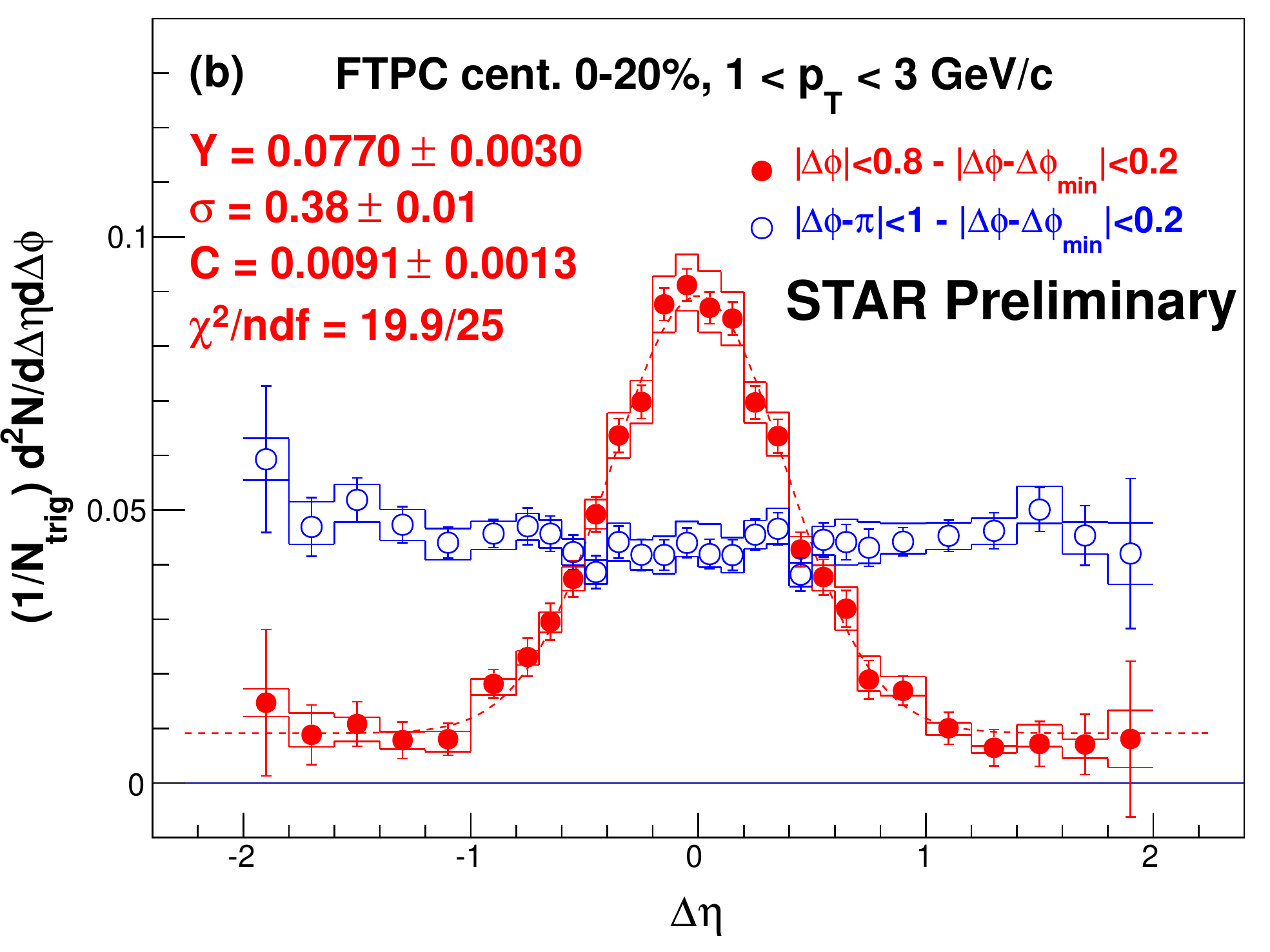}
\includegraphics[width=0.325\textwidth]{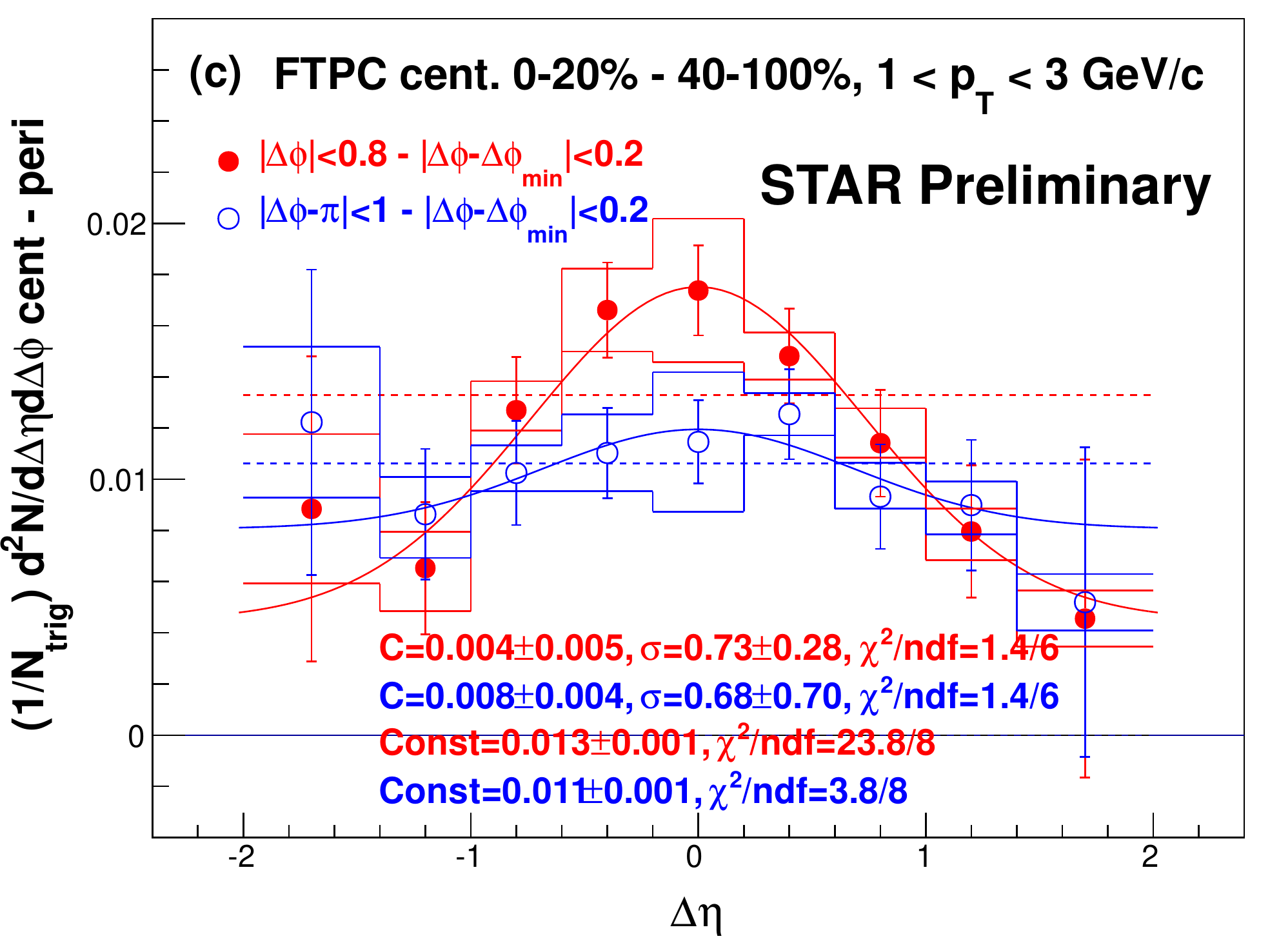}
\end{center}
\caption{Dihadron $\deta$ correlations for near side (red) and away side (blue) in peripheral (a) and central (b) \dau\ collisions, and their ``central $-$ peripheral'' differences (c). Trigger and associated particles are both from TPC ($|\eta|<1$) and $1<\pt<3$~\gev. Centrality is determined by FTPC-Au ($-3.8<\eta<-2.8$) multiplicity. Error bars are statistical and boxes indicate systematic uncertainties.}
\label{fig:deta}
\begin{center}
\includegraphics[width=0.245\textwidth]{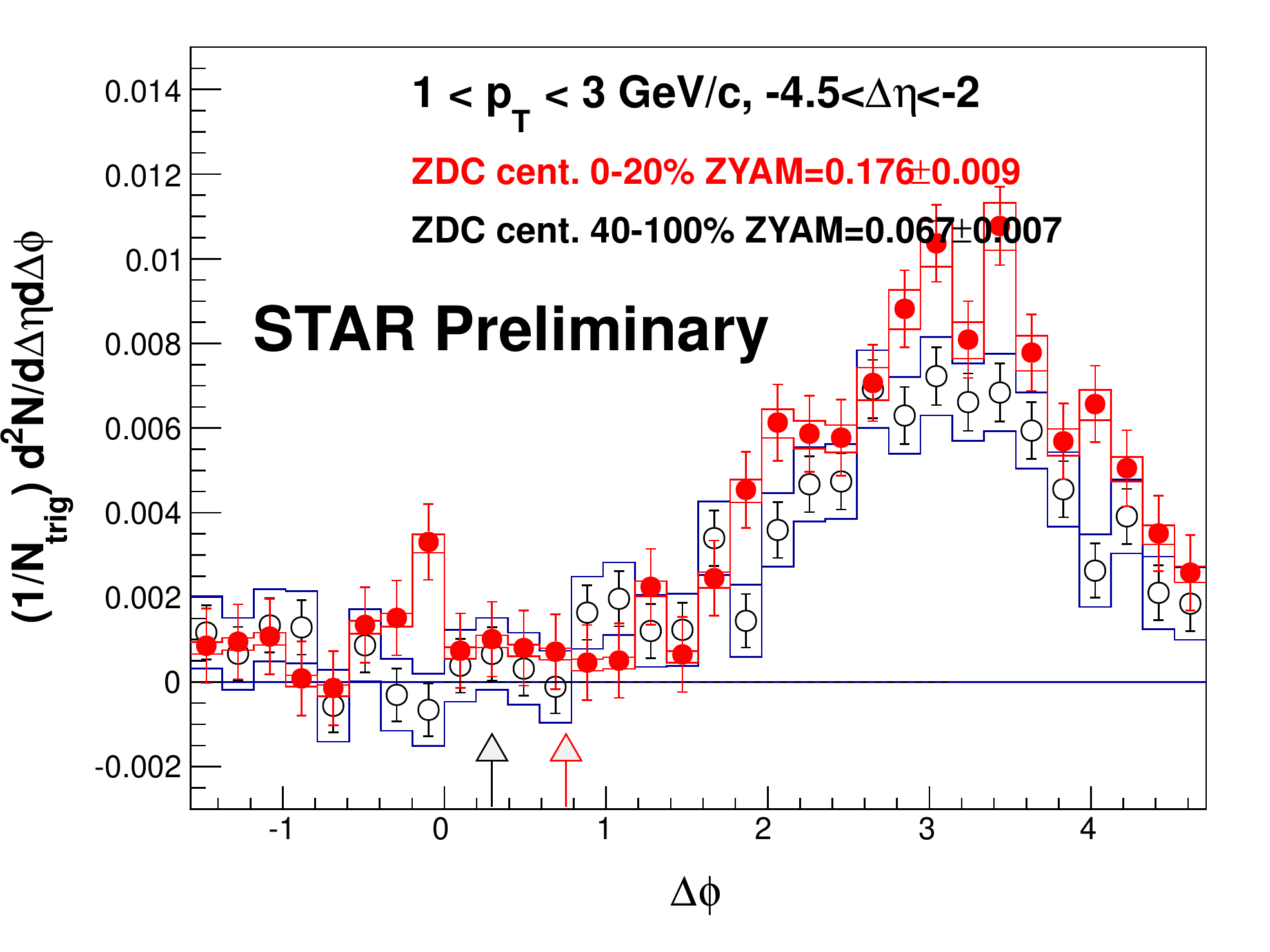}
\includegraphics[width=0.245\textwidth]{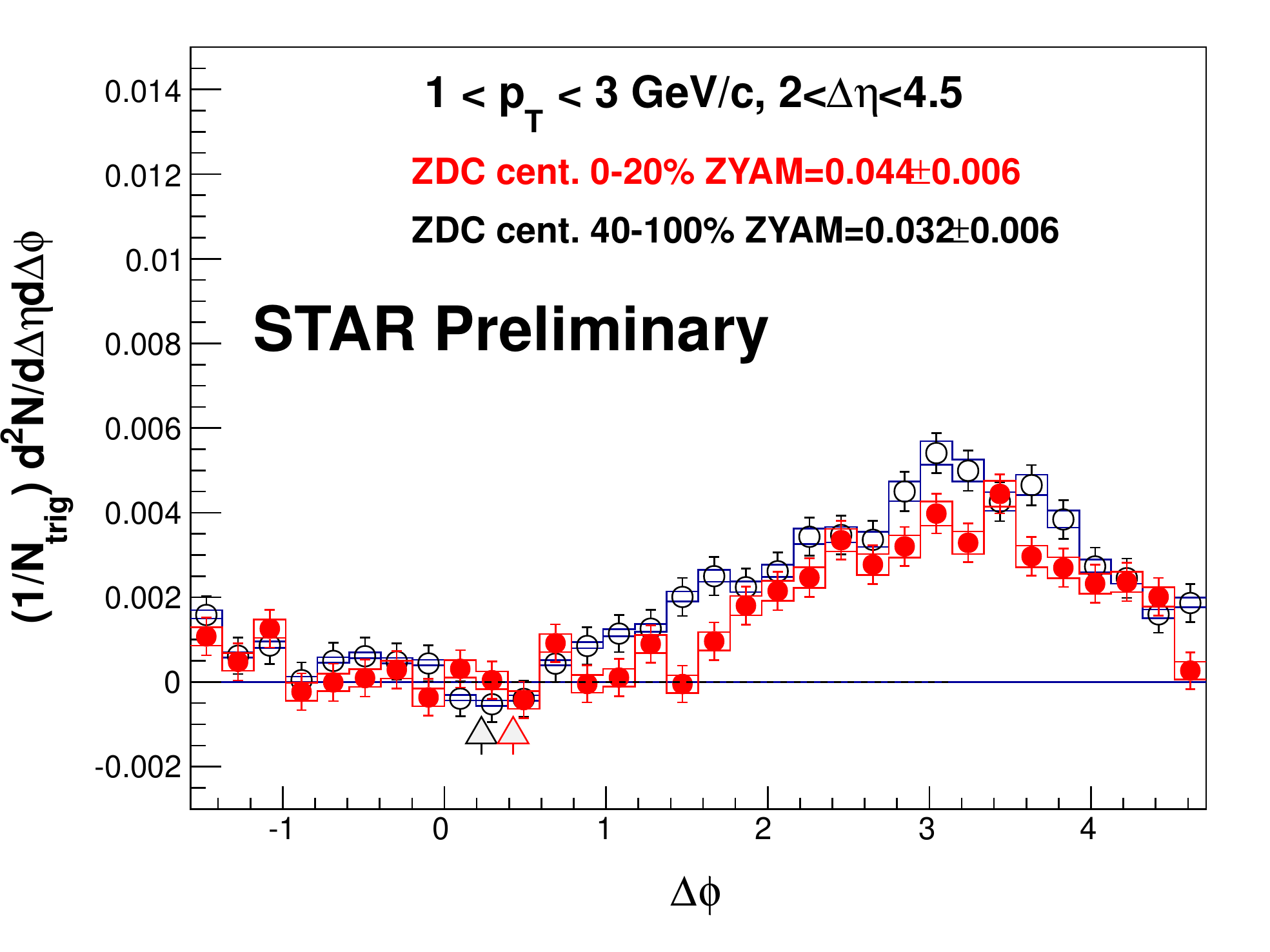}
\includegraphics[width=0.245\textwidth]{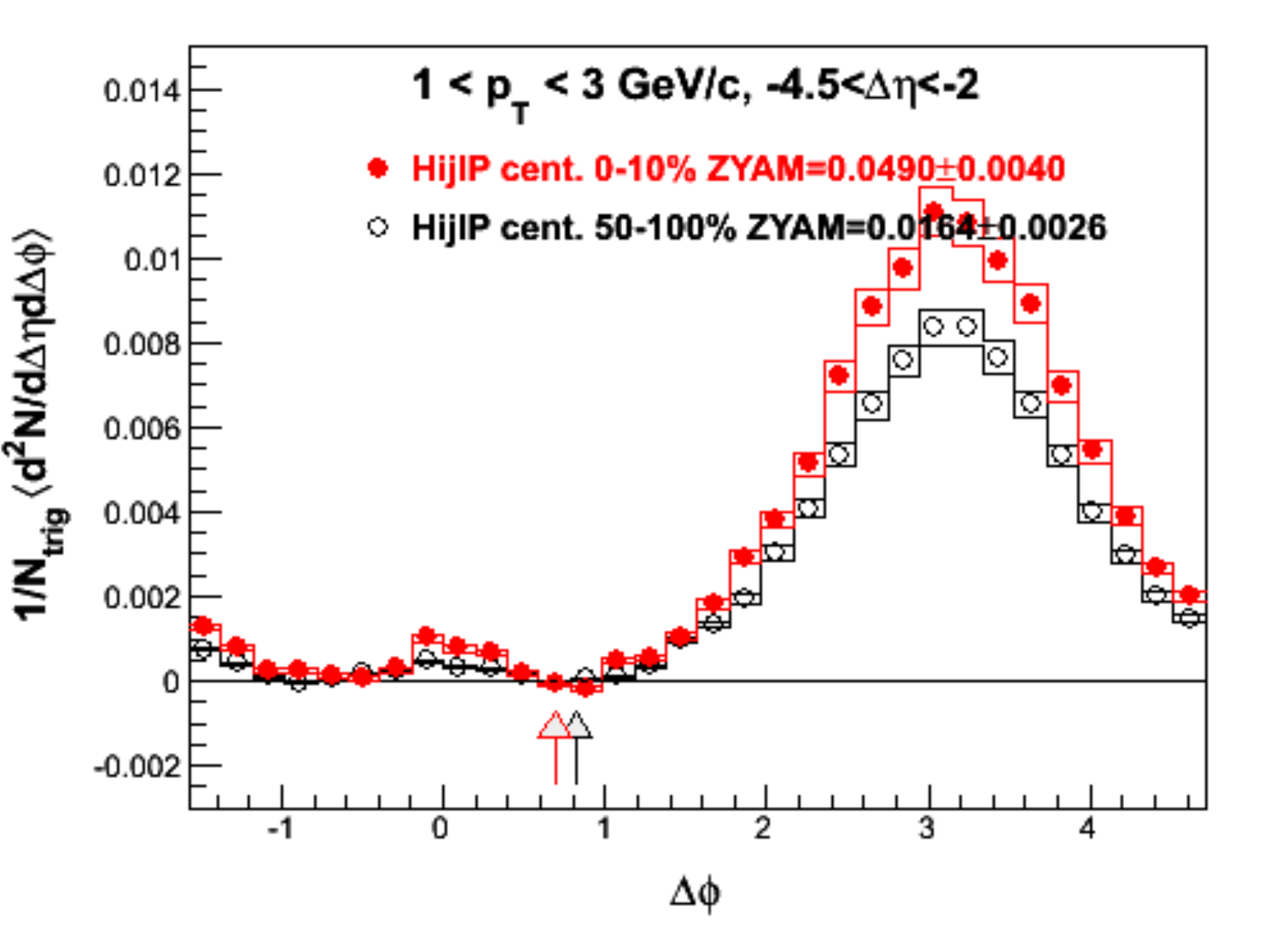}
\includegraphics[width=0.245\textwidth]{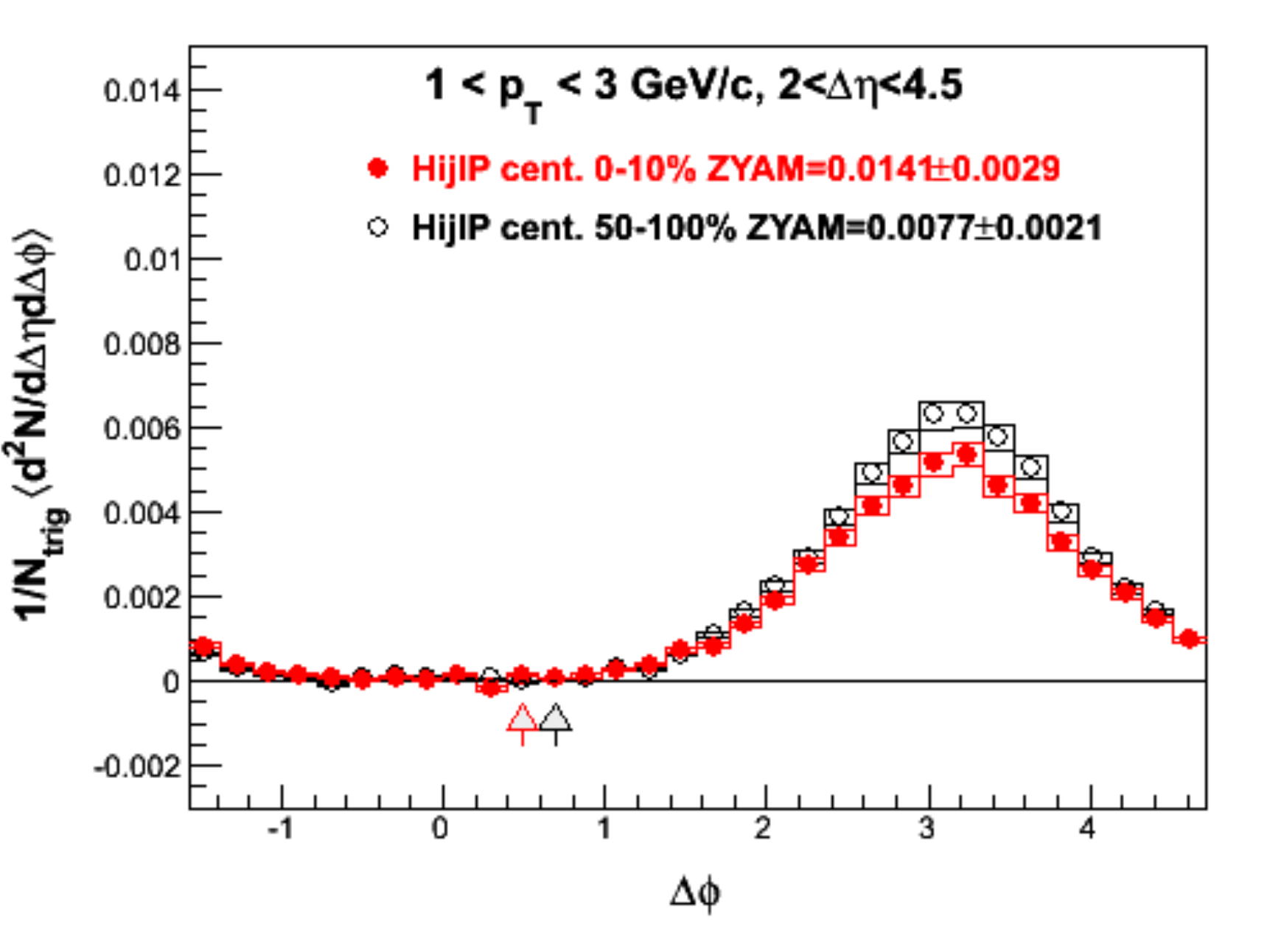}
\end{center}
\vspace{-1.3in}
\hspace{0.38in}{\footnotesize a)}\hspace{1.5in}{\footnotesize b)}
\hspace{1.48in}{\footnotesize c)}\hspace{1.5in}{\footnotesize d)}
\vspace{0.9in}
\caption{Dihadron $\dphi$ correlations at large $\deta$ in peripheral (black) and central (red) \dau\ collisions from data (a,b) and HIJING (c,d). Centrality is determined by ZDC-Au for data and impact parameter for HIJING. Trigger particles are from TPC ($|\eta|<1$); associated particles are from FTPC-Au (a,c) and FTPC-d (b,d). The arrows indicate ZYAM normalization locations. Error bars are statistical and boxes indicate systematic uncertainties.}
\label{fig:dphi_FTPC}
\begin{center}
\includegraphics[width=0.245\textwidth]{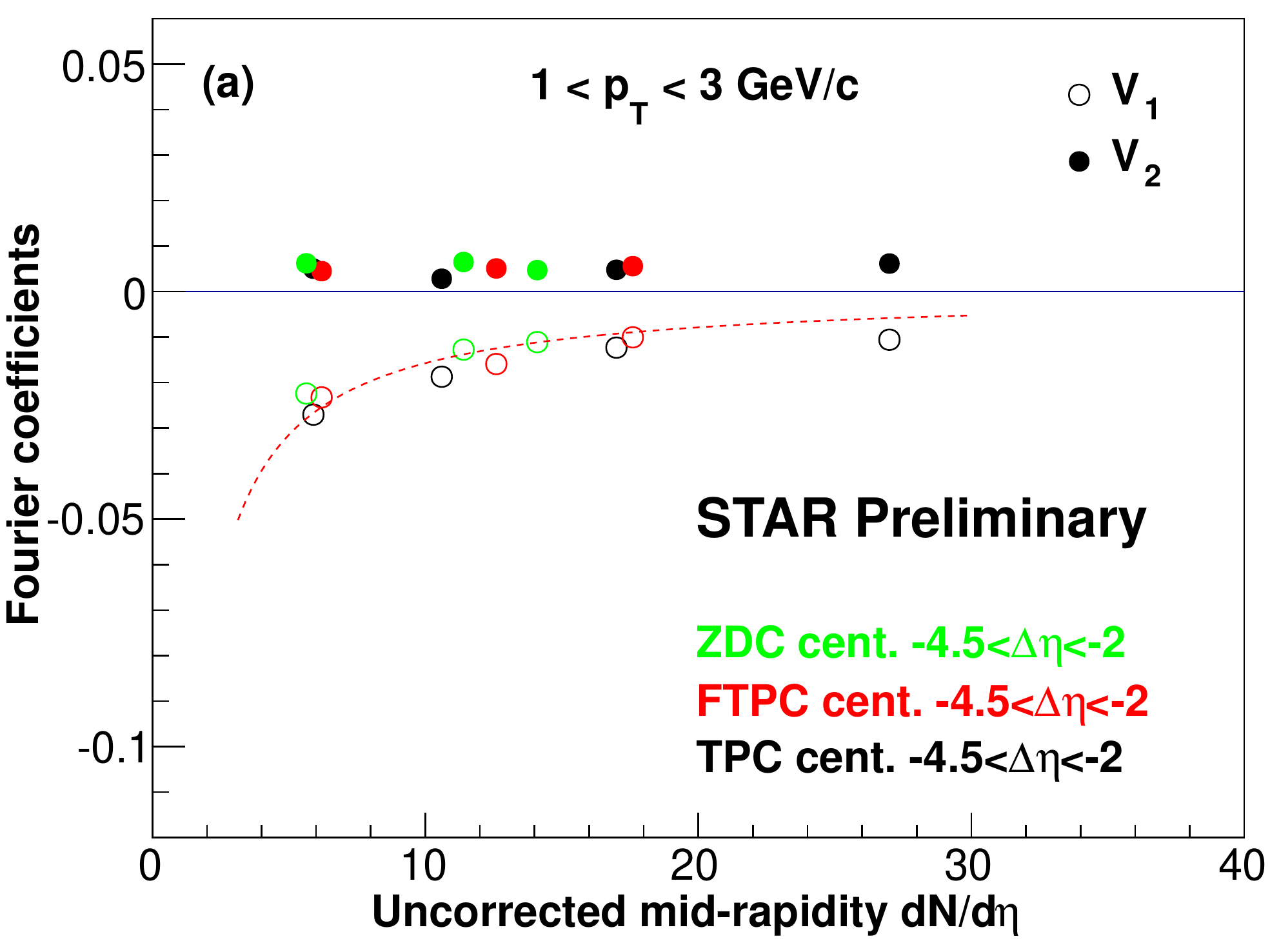}
\includegraphics[width=0.245\textwidth]{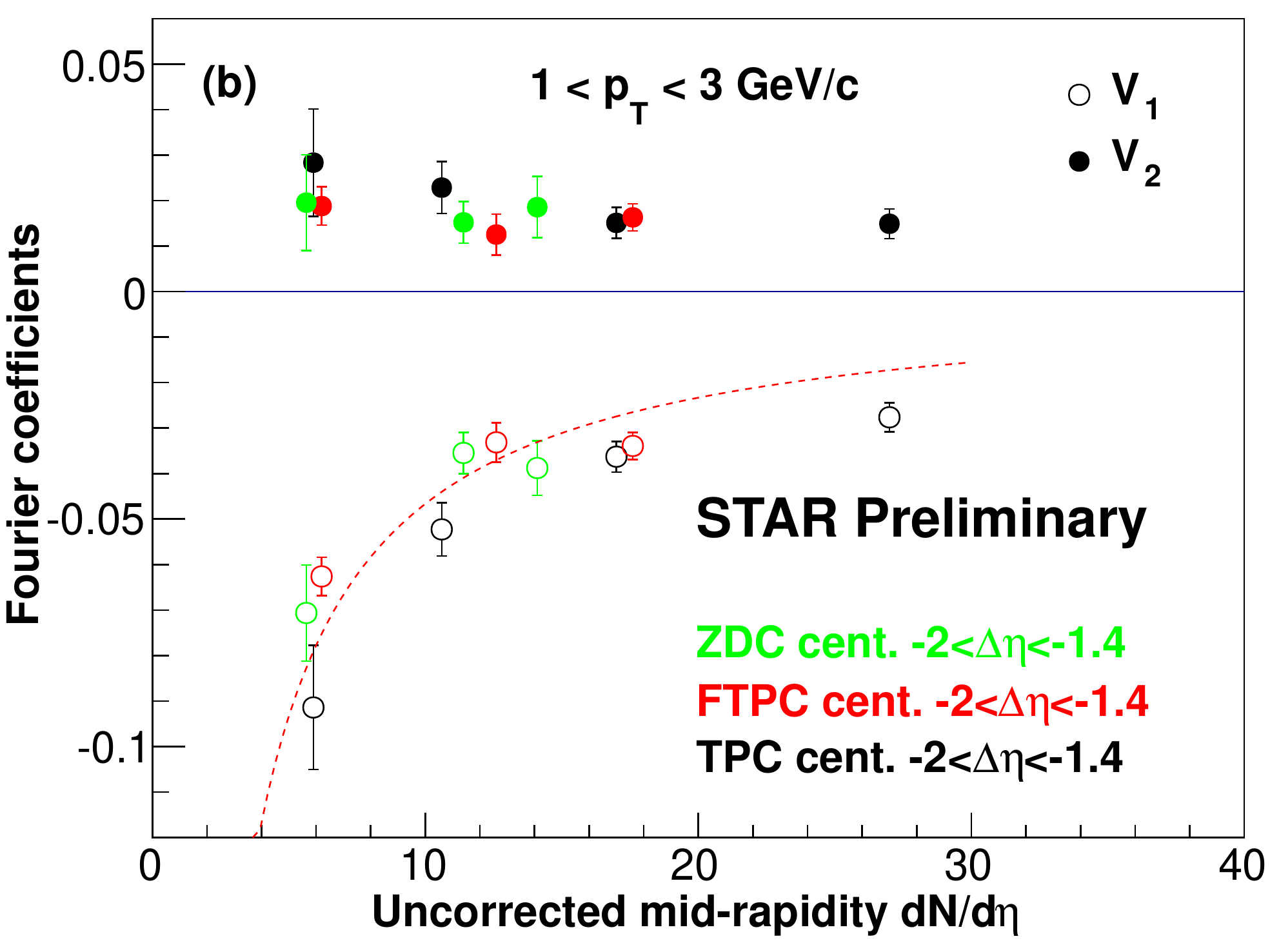}
\includegraphics[width=0.245\textwidth]{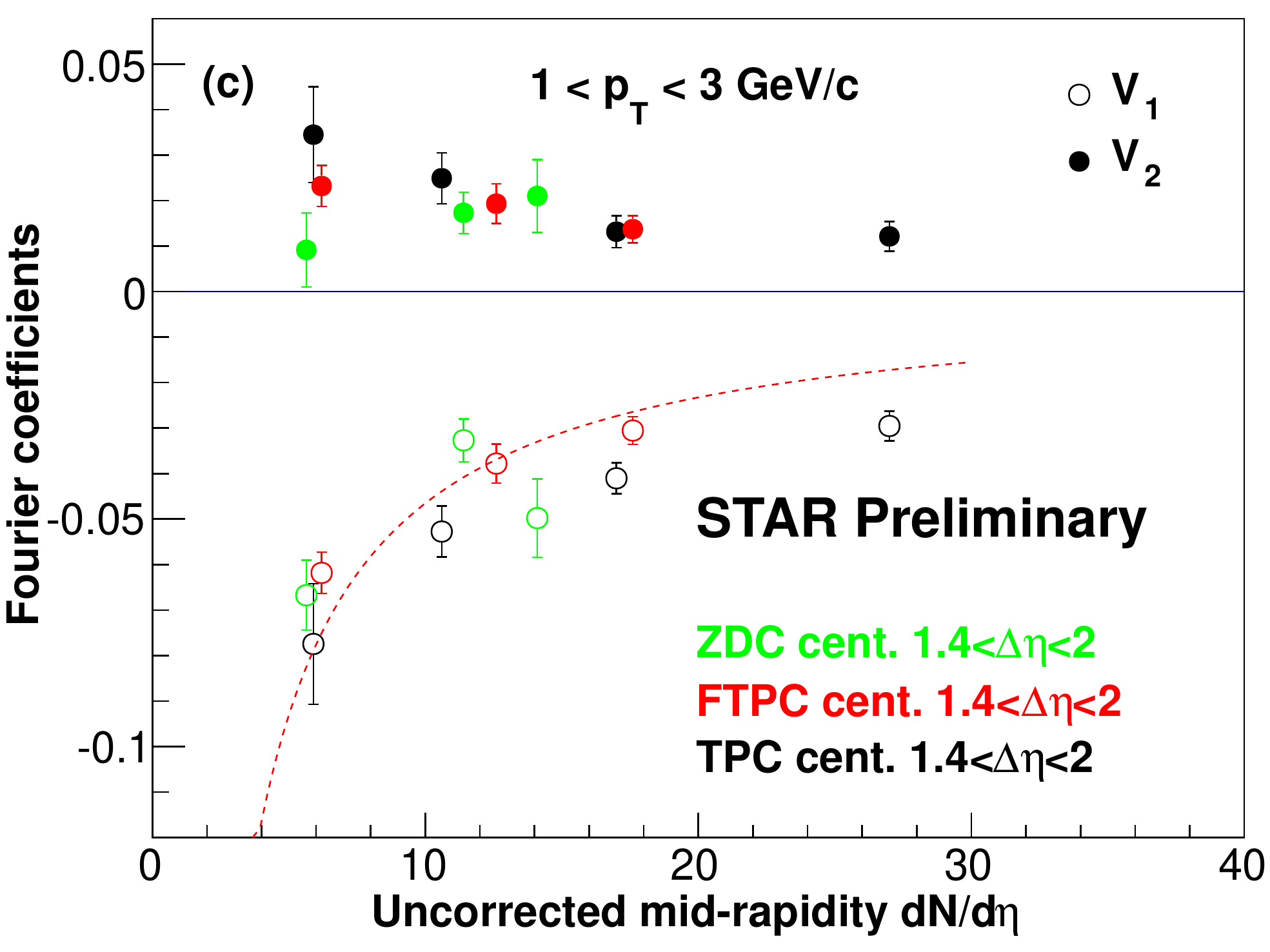}
\includegraphics[width=0.245\textwidth]{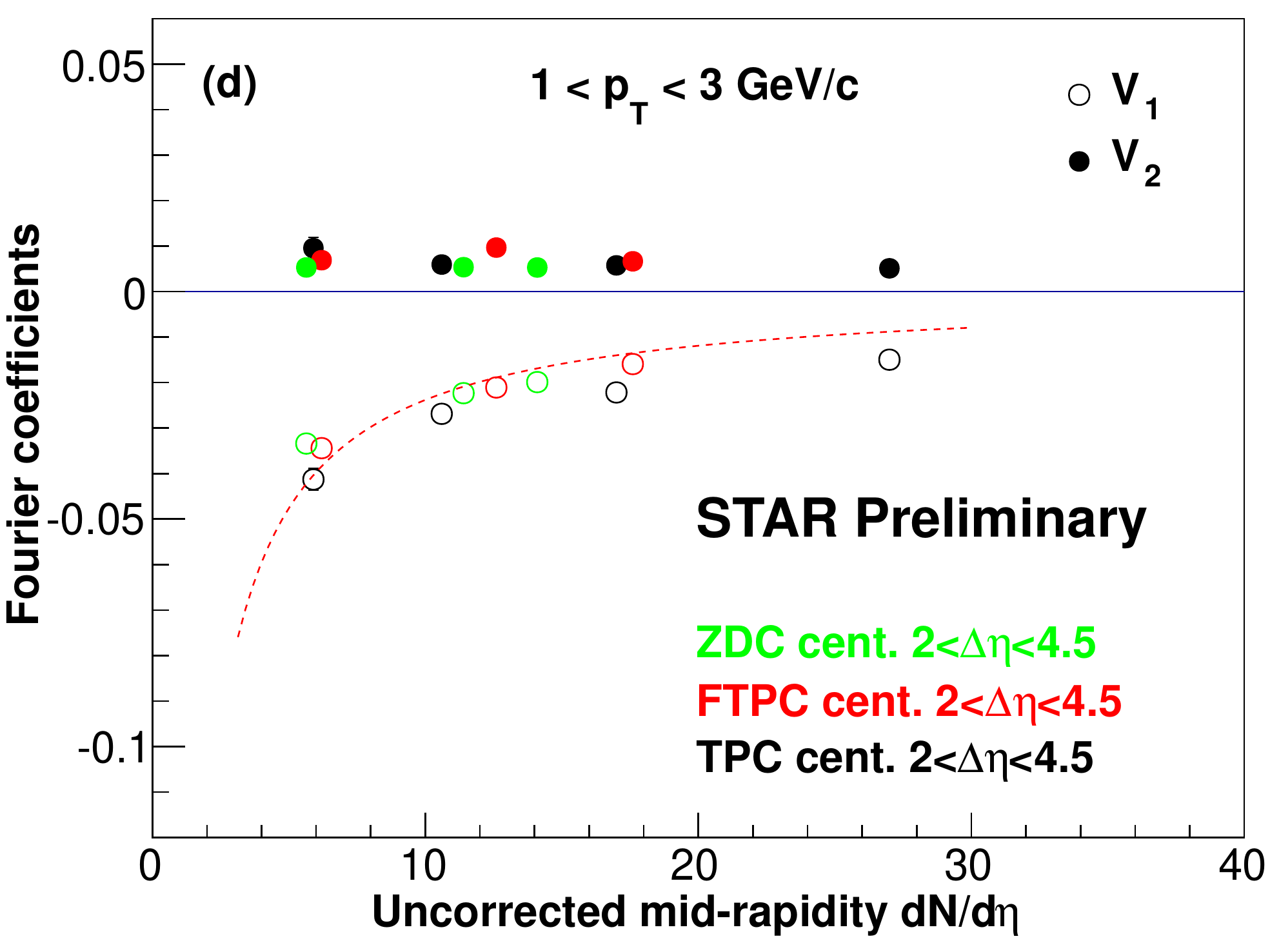}
\end{center}
\caption{Fourier coefficients of $\dphi$ correlation functions vs. event multiplicity in \dau\ collisions for four ranges of $\deta$. Results from all three centrality definitions are shown. Trigger and associated particle $\pt$ are both $1<\pt<3$~\gev. Error bars are statistical.}
\label{fig:VnVsMult}
\end{figure}

Figure~\ref{fig:dphi} shows the TPC-TPC $\dphi$ correlations in three ranges of $\deta$. Both peripheral and central collisions (determined by FTPC-Au) are shown. It is observed that the correlated yields are larger in central than peripheral \dau\ collisions. 

In order to investigate the source of the difference between central and peripheral collisions, $\deta$ correlations for near side ($|\dphi|<0.8$) and away side ($|\dphi-\pi|<1$) are shown in Fig.~\ref{fig:deta}(a,b). The near-side correlations exhibit a Gaussian peak and the away-side correlations are approximately uniform. Gaussian+constant fits to the near-side correlations indicate a difference of 20\% in the Gaussian area between central and peripheral collisions. The difference between central and peripheral collisions, shown in Fig.~\ref{fig:deta}(c), exhibits a near-side Gaussian peak and an approximate uniform away-side. These resemble the jet-correlation features, suggesting that the ``central $-$ peripheral'' difference is in these data mainly due to a difference in jet-like correlations. This difference is likely caused by biases in the centrality determination -- although FTPC-Au (used for centrality) is 3 units away from the correlation measurement, away-side jet-correlations can still contribute to the overall multiplicity in FTPC-Au. There can also be 
a physical centrality dependence, i.e.~cold nuclear matter effects.

TPC-FTPC dihadron correlations are shown in Fig.~\ref{fig:dphi_FTPC}(a,b) for associated particles from FTPC-Au and FTPC-d, respectively. On the Au-side, the away-side correlation is stronger in central than in peripheral collisions; on the d-side, it is the opposite. The same features are also observed in \dau\ collisions simulated by the HIJING model, shown in Fig.~\ref{fig:dphi_FTPC} (c,d). Surprisingly, there seem to exist finite correlated yields on the near side in the FTPC-Au acceptance far away from the trigger particle, in both data and HIJING. The underlying physics is unclear.


Figure~\ref{fig:VnVsMult} shows the Fourier coefficients, $V_n=\mean{\cos n\dphi}$ $(n=1,2)$. ($V_3$ is consistent with zero.) Four ranges of $\deta$ are shown, from left to right, for (a) TPC-FTPC-Au, (b) TPC-TPC with negative $\deta$, (c) TPC-TPC with positive $\deta$, and (d) TPC-FTPC-d correlations. Results with all three centrality determinations are shown, plotted at the corresponding measured mid-rapidity charged particle $dN/d\eta$. The $V_1$ is observed to approximately vary as $(dN/d\eta)^{-1}$, while the $V_2$ is approximately independent of $dN/d\eta$. $V_2$ is finite at all measured $\deta$; it is larger at mid-rapidity than forward/backward rapidities; $V_2$ from TPC-FTPC-d correlation may be even larger than that from TPC-FTPC-Au correlation.

In fact, the Fourier coefficients of the ``central $-$ peripheral'' difference distribution are no different from those of the peripheral and central collisions. 
This fact is governed by simple mathematics because the $V_n$ coefficients are taken relative to the mean of the distribution. However, the underlying physics mechanisms for the large Fourier coefficients of the ``central $-$ peripheral'' diffference are not entirely clear. Whether there are additional sources, except the aforementioned difference in jet-like correlations due to centrality biases, remains an open question. One of the future studies is to better quantify the centrality biases to jet-like correlations, and then investigate any additional physics mechanisms for the ``central $-$ peripheral'' difference. 

\section{Summary}
Dihadron $\dphi$ and $\deta$ correlations are reported for peripheral and central \dau\ collisions at $\snn=200$~GeV from STAR. The ZYAM background subtracted correlation yields at mid-rapidity are larger in central than peripheral collisions. The ``central $-$ peripheral'' differences resemble jet-like correlations, Gaussian peaked on the near side and approximately uniform on the away side. The difference is likely to be caused by a difference in jet-like correlations due to centrality biases. At forward rapidity, dihadron correlations on the away side are suppressed/enhanced in the d/Au-side direction of the more central d+Au collisions. Similar effects are present in the HIJING model.

Fourier coefficients of the raw dihadron correlations are also reported. The first harmonic coefficient is found to be approximately inversely proportional to event multiplicity. The second harmonic coefficient is found to decrease with $\deta$, but finite at forward/backward rapidity of $|\deta|\approx3$; it is approximately independent of the event multiplicity. 

The large acceptance of STAR allows detailed investigation of dihadron correlations and their centrality biases. The \dau\ data seem to be mainly consistent with jet phenomenology. The next step is to quantify ``central $-$ peripheral'' differences caused by centrality biases, and potentially isolate possible contributions unrelated to jets.

\section*{Acknowledgments}
This work was supported by U.S.~Department of Energy under Grant No. DE-FG02-88ER40412.


\begin{thebibliography}{10}
\expandafter\ifx\csname url\endcsname\relax
  \def\url#1{\texttt{#1}}\fi
\expandafter\ifx\csname urlprefix\endcsname\relax\def\urlprefix{URL }\fi
\expandafter\ifx\csname href\endcsname\relax
  \def\href#1#2{#2} \def\path#1{#1}\fi
\bibitem{CMSppRidge} V.~Khachatryan, et~al., JHEP 1009 (2010) 091.
\bibitem{CMSpPbRidge} S.~Chatrchyan, et~al., Phys.Lett. B718 (2013)  795.
\bibitem{ALICEpPbRidge} B.~Abelev, et~al., Phys.Lett. B719 (2013)  29; G.~Aad, et~al., Phys.Rev.Lett. 110 (2013) 182302.
\bibitem{PRL95} J.~Adams, et~al., Phys.Rev.Lett. 95 (2005) 152301; B.~Abelev, et~al., Phys.Rev. C80 (2009) 064912; B.~Alver, et~al., Phys.Rev.Lett. 104 (2010) 062301. 
\bibitem{AlverV3} B.~Alver, G.~Roland, Phys.Rev. C81 (2010) 054905, erratum-ibid.~{\bf C82},  039903 (2010).
\bibitem{Bozek:2010pb} P.~Bozek,  Eur.Phys.J. C71 (2011) 1530; P.~Bozek, W.~Broniowski, Phys.Lett. B718 (2013) 1557--1561.
\bibitem{Dumitru:2010iy} A.~Dumitru, et~al., Phys.Lett. B697 (2011) 21; K.~Dusling, R.~Venugopalan, Phys.Rev. D87  (2013) 054014; K.~Dusling, R.~Venugopalan, Phys.Rev. D87 (2013) 094034.
\bibitem{PHENIXdAuRidge} A.~Adare, et~al., Phys.Rev.Lett. 111 (2013) 212301.
\bibitem{STARflow05} J.~Adams, et~al., Phys.Rev. C72 (2005) 014904.
\bibitem{Levente09} B.~Abelev, et~al., Phys.Rev. C79 (2009) 034909.
\bibitem{ZYAM} N.~Ajitanand, et~al., Phys.Rev. C72 (2005) 011902.
\end{thebibliography}
\end{document}